\documentclass[prb,twocolumn,showpacs,preprintnumbers,amsmath,amssymb]{revtex4-1}

\usepackage{graphicx}% Include figure files
\usepackage{dcolumn}% Align table columns on decimal point
\usepackage{bm}% bold math
\usepackage{amssymb}%symbols of the figures
\usepackage{amsmath} 
\usepackage{wasysym}%other symbols of the figures
\usepackage{sidecap}
\usepackage{natbib} % BibTex related

\def\Ef{$E_{\rm F}$}

\def\kpara{\textit{{k}$_\parallel$}}

\def\invA{\AA$^{-1}$}

\def\Gbar{$\overline{\Gamma}$}

\def\GbarMbar{$\overline{\Gamma \rm M}$}%-$\overline{\rm M}$}
\def\GbarKbar{$\overline{\Gamma \rm K}$}%-$\overline{\rm K}$}
\def\BiTe{Bi$_2$Te$_3$}
\def\BiSe{Bi$_2$Se$_3$}

\DeclareMathOperator{\erf}{erf}

\begin{document}

\title{Observation of antiphase coherent phonons in the warped Dirac cone of \BiTe\ }

\author{E. Golias}
\email{evangelos.golias@gmail.com}
\author{J. S\'anchez-Barriga}
\email{jaime.sanchez-barriga@helmholtz-berlin.de}

\affiliation{Helmholtz-Zentrum Berlin f\"ur Materialien und Energie, Albert-Einstein Stra{\ss}e 15, 12489 Berlin, Germany}

\begin{abstract}

In this Rapid Communication we investigate the coupling between excited electrons and phonons in the highly anisotropic electronic structure of the prototypical topological insulator \BiTe. Using time- and angle-resolved photoemission spectroscopy we are able to identify the emergence and ultrafast temporal evolution of the longitudinal-optical A$_{1g}$ coherent phonon mode in \BiTe. We observe an anti-phase behavior in the onset of the coherent-phonon oscillations between the \GbarKbar\ and the \GbarMbar\ high-symmetry directions that is consistent with warping. The qualitative agreement between our density-functional theory calculations and the experimental results reveals the critical role of the anisotropic coupling between Dirac fermions and phonon modes in the topological insulator \BiTe.
 
\end{abstract}


\maketitle

%%%    Introduction    %%%
Topological insulators (TIs) are novel materials with bulk insulating behavior and robust spin-polarized metallic states at the surface as imposed by the principle of bulk-boundary correspondence. \cite{Hasan:2010ku, Qi:2011hb} Topological surface states (TSSs) in TIs, protected by the time-reversal symmetry, forbid backscattering properties and support dissipationless pure spin currents that can be utilized in future spintronic devices. \cite{Ando:2013kb} Angle-resolved photoemission spectroscopy (ARPES) has been used as the main experimental technique for the discovery of novel topological phases, \cite{Hsieh2009sc} as together with spin resolution, \cite{Hsieh2009sc,Sanchez-Barriga-PRX-2014} it gives direct access to the electronic structure of materials with unprecedented detail.\cite{Hufner:1996bok} 

During the past years, considerable effort has been focused on the study of the so-called prototypical three-dimensional TIs, namely Bi$_2$Se$_3$, Sb$_2$Te$_3$ and Bi$_2$Te$_3$, where TSSs have been theoretically predicted \cite{Zhang:2009ks} and observed experimentally using ARPES. \cite{Chen:2009er, Xia:2009fn, Hsieh:2009ig} Furthermore, photoemission has been used for the investigation of the underlying mechanisms governing the scattering processes of Dirac fermions in TSSs \cite{Valla:2012sc, Barriga:2014ws} as well as their coupling to bosonic modes in real TI materials. \cite{Park:2011fq, Pan:2012wr, Kondo:2013fo} The interactions between collective modes - especially phonons -  and Dirac fermions are of critical importance for potential device applications exploiting the TSS carriers, as in operating conditions transport is limited by electron-phonon scattering events in the non-ballistic regime of modern miniaturized transistors.

Coherent phonons can be easily triggered by optical excitation, hence, ultrashort laser pulses have been routinely employed in the study of the generation and dynamics of bosonic modes in solid materials. \cite{Bovensieper:2010bb} The introduction of a time delay between two ultrashort optical pulses, namely, the pump and probe pulse, facilitates the investigation of coherent phonon oscillations in the time domain. This methodology has been extensively employed by different spectroscopic methods for the study of various out-of-equilibrium phenomena.\cite{Dekorsy:2000fm}

More specifically, numerous time-resolved spectroscopic studies have established the observation of coherent phonons in TIs following a perturbation by intense-laser fields,\cite{Wu:2008iv, Wang:2013fl, Flock:2014bw, Misochko:2015bp, Bykov:2015fa, kumar:2011rt} however, the subtle details of the momentum-dependent coupling between coherent phonons and TSSs have not been elucidated so far, mainly due to the utilization of pump-probe techniques that lack momentum resolution. The combination of pump-probe schemes and ARPES, namely time-resolved ARPES (tr-ARPES), can circumvent this deficiency by combining the access to the quasiparticle properties in the frequency domain and the information extracted from the collective modes in the time domain.\cite{Schmitt:2008jg} 

Most recently, tr-ARPES experiments on the prototypical TI \BiSe\ revealed that surface and bulk electrons couple differently to coherent phonons after optical excitation.\cite{Sobota:2014gu} Such an important observation opens the way for the investigation of other intriguing phenomena pertinent to TSSs that can only be accessed by combining time, energy and momentum resolution in a single experiment. 

In particular, tr-ARPES can be employed to investigate potential implications of anisotropies in the electronic structure of TIs for the generation and temporal evolution of coherent-phonon excitations coupled to TSSs. For such an investigation, \BiTe\ exemplifies the ideal candidate material due to its renowned anisotropic electronic structure, which manifests in the warped shape of its Fermi surface.\cite{Fu:2009ey} It has been predicted that the warping of the Fermi surface would have important effects in spectroscopic experiments.\cite{Fu:2009ey} In fact, by utilizing ARPES without time resolution it has been shown that under equilibrium conditions the warped Dirac cone of \BiTe\ underlies the anisotropic behavior of the photohole scattering times.\cite{Barriga:2014ws} Therefore, the interplay between phonons and Dirac fermions in the anisotropic Fermi surface of \BiTe\ out-of-equilibrium conditions should be also placed under scrutiny to investigate emergent phenomena that require the full potential of tr-ARPES.

In this Rapid Communication, we employ tr-ARPES to examine the generation of the longitudinal-optical coherent A$_{1g}$ phonon in the prototypical TI \BiTe\ and its coupling to the TSS. We are able to experimentally resolve an anti-phase in the onset of the coherent excitation along the different high-symmetry directions of the surface Brillouin zone (SBZ), which is remarkably reproduced by our \textit{ab initio} calculations within the framework of density functional theory (DFT).

%%%    Methods    %%%
The tr-ARPES experiments were performed at room temperature in an ultrahigh vacuum chamber with a base pressure of better than $1\cdot 10^{-10}$ mbar, where \BiTe\ single crystals, grown by the Bridgman method, were cleaved {\it in situ}. The excited electrons were collected by a Scienta R8000 hemispherical electron analyzer with an energy and angular resolution of 30 meV and 0.3$^{\circ}$, respectively. As pump and probe pulses we used the first (1.5 eV) and fourth (6 eV) harmonics of a homemade fs-laser system coupled to an ultrafast amplifier operating at 150 kHz. The light impinged the sample with an angle of 45$^{\circ}$ with respect to the surface normal, and the linear polarization of the pump and probe pulses was oriented parallel to both the analyzer slit and the sample surface. The time-resolution of the experiment was $\sim$ 200 fs and the pump fluence $\sim$ 300 $\mu$J/cm$^2$. 

The DFT calculations were performed using plane-wave potentials \cite{Blochl:1994paw} and the exchange-correlation functional of Perdew, Burke and Ernzerhof \cite{pbe:1996pp, *pbe:1996er} as implemented in the VASP package. \cite{kresse:1999vas} The long-range dispersive forces were taken into account utilizing the DFT-D3 methodology of Grimme \cite{grimme:2010dm} and the spin-orbit interaction was included using the scalar-relativistic approximation. The electronic orbitals were expanded in a plane-wave basis up to kinetic energies of 300 eV and the reciprocal space of the basal plane was sampled with a $\Gamma$-centered 11$\times$11 k-point grid. 

\begin{figure}
\includegraphics [width=0.48\textwidth]{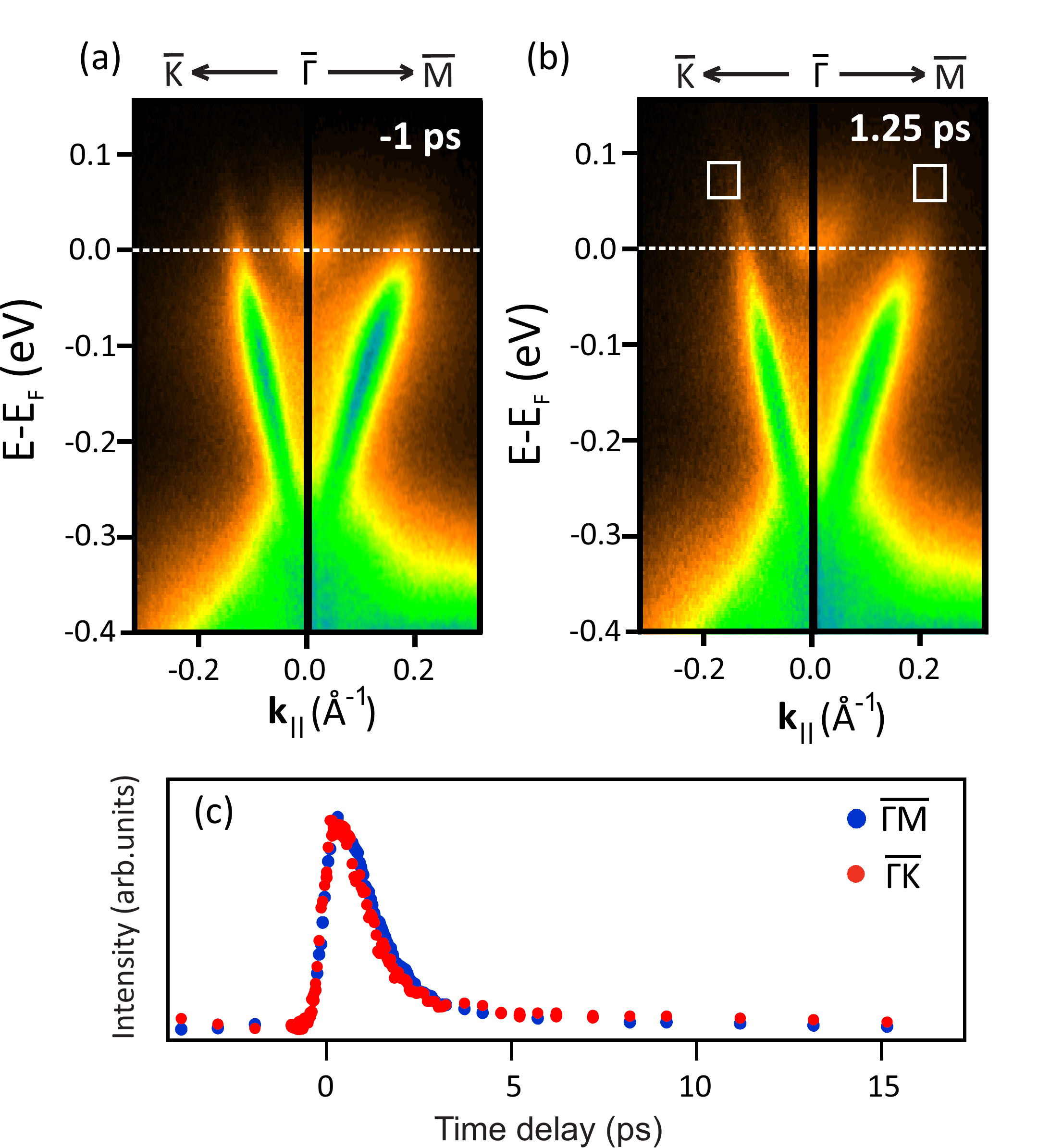}
\caption{High-resolution tr-ARPES spectra along the \GbarKbar\ and \GbarMbar\ high-symmetry directions at pump-probe delays of (a) -1 ps and (b) 1.25 ps. The white-dashed line indicates the position of the Fermi level. (c) Corresponding tr-ARPES intensities along \GbarMbar\ (blue) and \GbarKbar\ (red) integrated in the areas above the Fermi level marked with white boxes in (b).}
\label{fig1}
\end{figure}

%%%  Main Text %%%

Figure \ref{fig1} presents tr-ARPES spectra taken along the two nonequivalent \GbarKbar\ and \GbarMbar\ high-symmetry directions of the \BiTe\ SBZ.\cite{Supplement} In Fig. \ref{fig1}(a), where the time delay is negative, we probe the electronic structure in equilibrium. The divergence from the linear dispersion of the TSS along the \GbarMbar\ direction is in agreement with static ARPES measurements.\cite{Barriga:2014ws} Moreover, due to the $n$-doping, we can observe the bulk conduction band (BCB) with a parabolic dispersion crossing the Fermi level (\Ef). The pump pulse excites a transient electron population above \Ef\ that subsequently thermalizes within $\sim$ 500 fs due to the delayed electron filling from higher-energy states.\cite{Supplement} Figure \ref{fig1}(c) displays the temporal evolution of the tr-ARPES intensities integrated in the energy-momentum windows marked in Fig. \ref{fig1}(b) along the two nonequivalent high-symmetry directions. Fitting the temporal evolution of the integrated intensity to a function proportional to $(1+\erf(\frac{t}{\sigma}-\frac{\sigma}{\tau})) e^{-\frac{t-t_0}{\tau}}$, where  $\tau$ is the decay constant and $\sigma$ is the standard deviation of the Gaussian that corresponds to the time resolution of the experiment, we obtain time constants of 1.22 $\pm$ 0.02 ps and 1.16 $\pm$ 0.03 ps for \GbarKbar\ and \GbarMbar\ directions, respectively. Thus, the anisotropy of the Fermi surface has virtually no effect on the relaxation of the transient hot-electron population, which is consistent with previous tr-ARPES studies on \BiTe.\cite{Hajlaoui:2012cw, Hajlaoui:2013bx} In Fig. \ref{fig1}(c) it can also be seen that the hot electrons have almost completely decayed after $\sim$ 3 ps, which corresponds to the timescale where the coupling between hot electrons and lattice atoms through phonons drive the system towards equilibrium.

 \begin{figure}
 \includegraphics [width=0.48\textwidth]{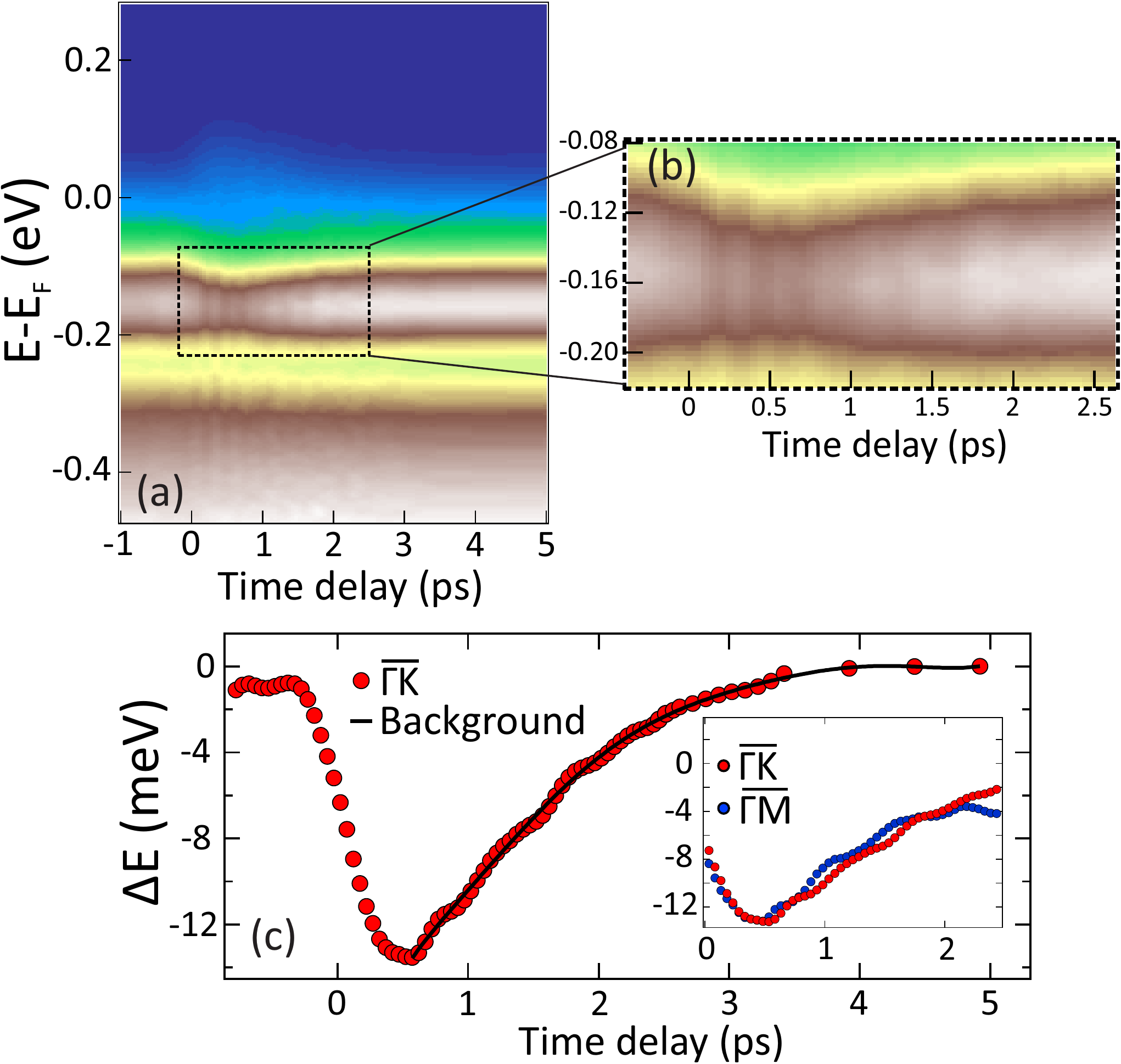}
 \caption{(a) Time evolution of the spectral weight at \kpara\ = 0.07 \invA\ along the \GbarKbar\ direction. (b) Zoom-in on the dashed region marked in (a), where we observe the period energy shift of the TSS as ripples in the time and energy domains. (c) Energy shift of the Lorentzian peak position extracted from the fitted EDCs of (a). Red circles depict the relative energy positions for different pump-probe delays, and the black-solid line corresponds to the incoherent background. Inset: Close comparison between the oscillations along \GbarKbar\ and \GbarMbar.}
 \label{fig2}
 \end{figure}

In Fig. \ref{fig2}, we focus on the dynamics of the occupied part of the TSS. Figure \ref{fig2}(a) illustrates the temporal evolution of the TSS at \kpara = 0.07 \invA and 150 meV below \Ef\ along \GbarKbar. After the arrival of the pump pulse, we observe a suppression of the TSS intensity as a result of the hole creation due to the optical excitation. In Fig. \ref{fig2}(b), we zoom-in on the energy-time region where the TSS is recovering from the initial optical excitation, and observe a ripple-like shape at the border of the spectral intensity of the TSS. Fitting a Lorentzian curve convoluted with a Gaussian, which represents the energy resolution of our experiment, to the energy-distribution curves (EDCs) of Fig. 2(a) and extracting the position of the peak with respect to the unperturbed spectrum,\cite{Supplement} leads to the binding energy (BE) shift displayed in Fig \ref{fig2}(c). Initially, the optical pump pulse shifts the TSS towards higher BE, a photoinduced effect that has been also observed in the metallic surface states of Bi (Ref.~\onlinecite{Papalazarou:2012hy}) and Gd. \cite{Bovensiepen:2007ug} After the initial sharp energy shift upon optical excitation the TSS recovers towards its initial BE position within $\sim$ 3 ps, a process which is accompanied by the repopulation of the TSS. From Fig. \ref{fig2}(c) it is evident that the recovery of the TSS encompasses a coherent component, also seen along the \GbarMbar\ direction in the inset, which oscillates around the incoherent part of the curves. We isolate the coherent oscillations by subtracting a smooth function, a tenth degree polynomial,\cite{Sobota:2014gu} from the corresponding curves. The same analysis procedure was used for the TSS along the \GbarMbar\ direction and for the BCB near the \Gbar\ point at a BE of $\sim$100 meV and $\sim$20 meV, respectively.\cite{Supplement}

\begin{figure}
 \includegraphics [width=0.45\textwidth]{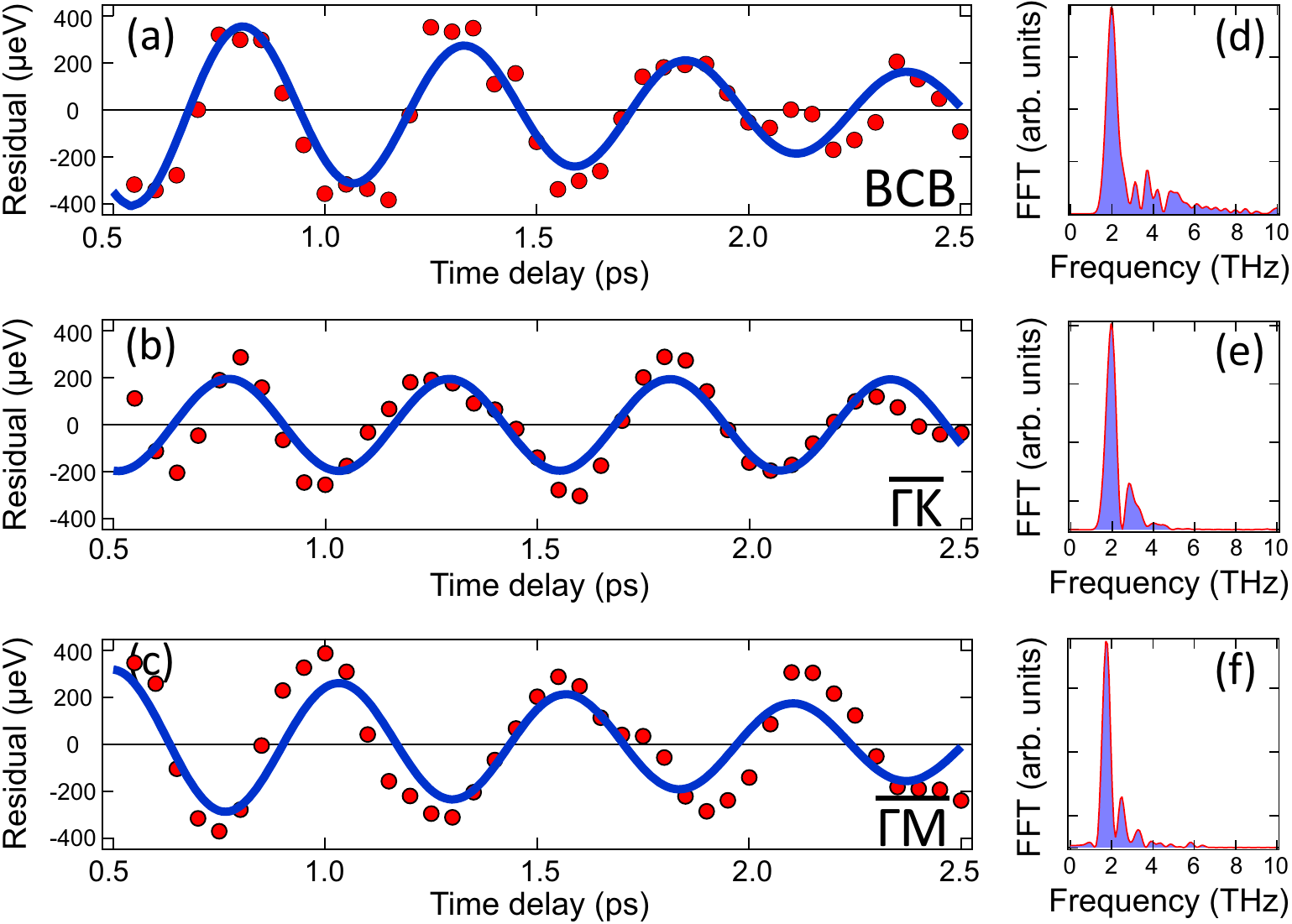}
 \caption{Residuals after subtracting the incoherent background in the energy positions of (a) the BCB near \Gbar, and the TSS along (b) \GbarKbar\ and (c) \GbarMbar\ directions, respectively. Red circles represent the calculated residuals while blue lines are the best fit of a periodic exponential decay function. (d)–(f) Squared magnitude of the Fourier transform of the experimental data in (a)–(c), respectively.}
 \label{fig3}
 \end{figure}

In Fig. \ref{fig3}(a)-(c) we observe the oscillatory components of the BCB and TSS as measured along \GbarKbar\ and \GbarMbar. In Fig. \ref{fig3}(d)-(f), we plot the square amplitude of the Fourier transform of the experimental data in order to extract the dominant periodic component of the oscillatory profiles. The principal frequencies for the analyzed regions are in excellent agreement with the frequency of the zone-centered longitudinal-optical A$_{1g}$ phonon in \BiTe.\cite{Kullmann:1984, Wu:2008iv, Wang:2013fl, Flock:2014bw, Weis:2015hb, Misochko:2015bp} More specifically, by fitting a periodic decaying function of the form $A\cos(2\pi f t+\phi)e^{-\frac{t}{\tau}}$ to the data of Figs. \ref{fig3}(a)-(c) we obtain the oscillation frequencies of f$_1$ = 1.91 $\pm$ 0.04 THz, f$_2$ = 1.92$\pm$0.03 THz and f$_3$ = 1.87$\pm$0.04 THz for the BCB and the TSS along the \GbarKbar\ and \GbarMbar\ directions, respectively. It is interesting to note that the onset of the coherent phonon finds the oscillations in Figs. 3(a)-(c) at their extreme positions, in consequence following a cosine-like profile. This behavior pinpoints a process known as displacive excitation of coherent phonons (DECP)\cite{Zeiger:1992de} as the underlying mechanism that drives the coherent-phonon oscillations in \BiTe. We emphasize that the DECP mechanism would require a zero initial phase $\phi$ at t = 0. In our experiment the oscillations start at t = 0.5 ps, thus the extrapolated zero initial phase cannot be used as the relevant quantity to directly justify the DECP mechanism based on the original model proposed by Zeiger {\it et al}.\cite{Zeiger:1992de} However, the DECP mechanism is further supported by the fact that the pulse duration in our experiment is smaller than the period of the phonon,\cite{Zeiger:1992de, Nakamura} thus providing an impulsive force that drives the atoms towards their new equilibrium positions with zero kinetic energy. This process is typically associated with the light-electron interaction in opaque materials such as \BiTe\ and it is accompanied by the excitation of the fully-symmetric A$_{1g}$ phonon mode. More specifically, it has been shown that under similar experimental conditions the principal driving force of the A$_{1g}$ coherent phonon in \BiTe\ is the temperature gradient due to the transient hot-electron population,\cite{Wang:2013fl} which explains the excitation of the A$_{1g}$ phonon in our experiment despite the fact that our light polarization is fully in-plane. Here, the most intriguing characteristic of the coupling between electrons and phonons is the $\pi$ phase difference that we observe between the warped and the linear branches of the TSS during the onset of the coherent oscillation [compare Figs. 3(b) and 3(c)]. This behavior strongly indicates that the anti-phase oscillation originates from the anisotropic electronic structure of \BiTe\ and emerges due to the coupling with the hot-electron transient as an anti-phase in the onset of the coherent-phonon oscillation. 

\begin{figure}
\includegraphics [width=0.42\textwidth]{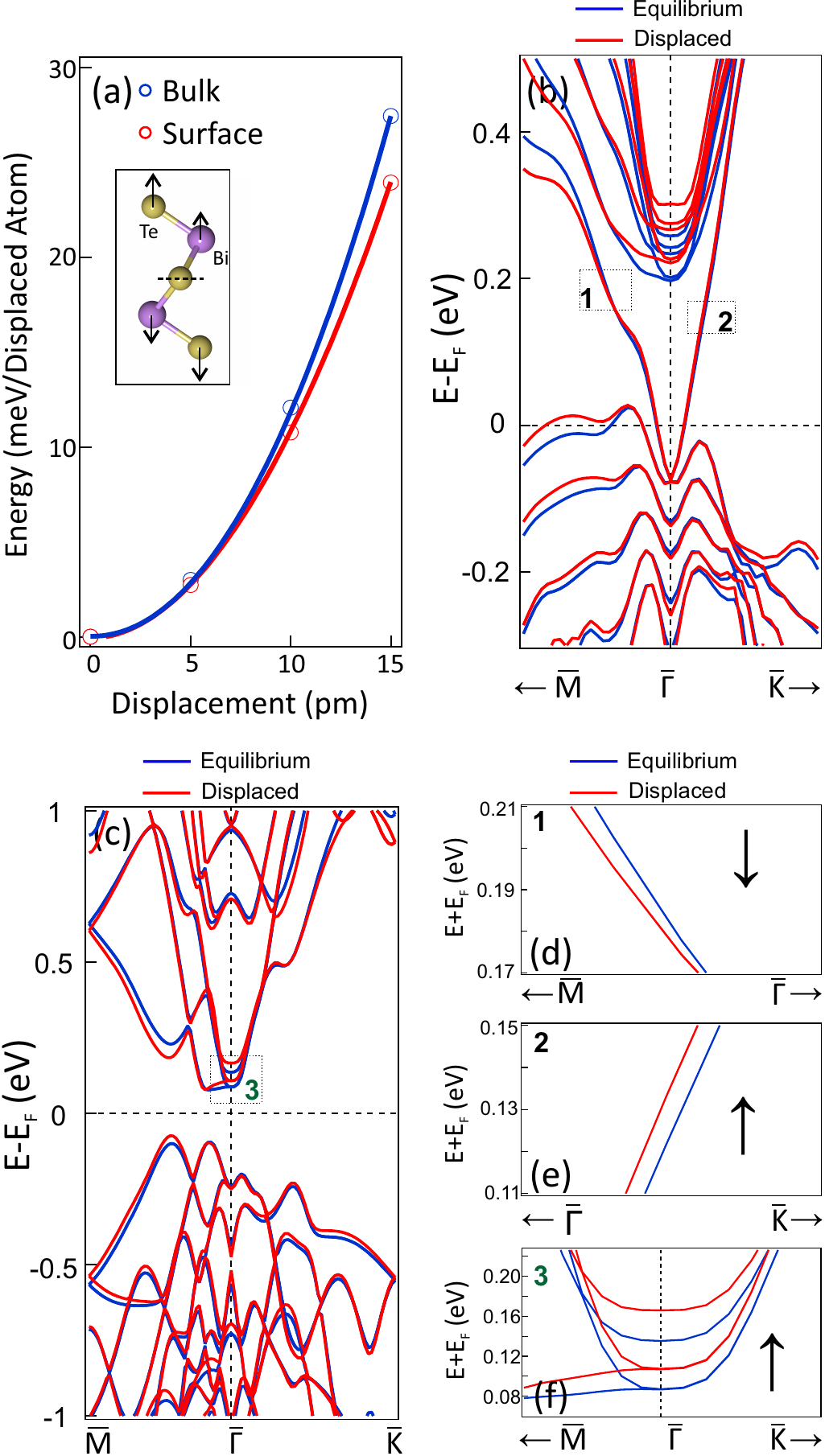}
\caption{(a) Energy difference of the frozen-phonon calculation with respect to the ground-state energy of the bulk (blue circles) and slab (red circles) along with their corresponding parabolic fits (solid lines). In the inset, black arrows depict the applied distortions in the QLs of \BiTe\ in order to simulate the A$_{1g}$ phonon. Yellow and purple spheres represent Te and Bi atoms, respectively. (b) Slab and (c) bulk band structure calculation where blue and red lines represent the bands in equilibrium and in a 5 pm distorted configuration, respectively. (d)-(f) Zoom-in on the numbered regions marked with a dashed rectangle in (b) and (c). Black arrows indicate the direction of the band displacement with respect to the equilibrium band structure.}
\label{fig4}
\end{figure}

In order to shed light on the origin of the observed anti-phase we employed \textit{ab initio} DFT calculations. We used a hexagonal cell comprising six quintuple layers (QLs) for the bulk and six QLs for the slab configuration with more than 40 \AA\ of vacuum separating the opposite surfaces. For both frozen-phonon calculations we displaced the atoms in every QL as shown in the inset of Fig. \ref{fig4}(a), in accordance with the symmetry of the A$_{1g}$ phonon mode. Figure \ref{fig4}(a) shows the energy difference with respect to the equilibrium calculation for the bulk and slab configurations, where we observe that small displacements ($\lesssim$ 5 pm) lead to insignificant energy differences between the bulk and surface systems. This is in agreement with our observation that bulk and surface electrons exhibit similar frequencies, a result that correlates well with the fact that the previously observed softening of the A$_{1g}$ phonon mode in \BiSe\ \cite{Sobota:2014gu} does not exist in the present case due to the lower pump fluence used in our study.
 
In Fig. \ref{fig4}(b) we illustrate the band structure of the \BiTe\ slab, which displays an anisotropic TSS in agreement with previous DFT calculations, \cite{Park:2010ia, Michiardi:2014dw} while in Fig. \ref{fig4}(c) we plot the calculated bulk band structure. In Figs. \ref{fig4}(b) and 4(c), where the equilibrium and frozen-phonon band structures are juxtaposed, we mark three different regions where we probed the coherent-phonon oscillations in our tr-ARPES experiment. Figures \ref{fig4}(d)-\ref{fig4}(f) display a focused plot in the aforementioned energy-momentum windows, where the anti-phase observed in our tr-ARPES data is fully reproduced by our calculations. We can see that when the atoms are displaced according to the A$_{1g}$ phonon mode, the warped branch of the TSS along \GbarMbar\ moves downwards, while the linear branch along \GbarKbar\ as well as the BCB move in the opposite direction. This observation strongly suggests that the origin of the anti-phase oscillation is the surface-state warping, which is ultimately connected to the level repulsion between the surface and bulk bands bending down along the \GbarMbar\ direction.\cite{Fu:2009ey} Finally, we note that the calculations in Fig. 4(c) predict the bulk-valence band (BVB) to exhibit anti-phase oscillations with respect to the BCB near the \Gbar\ point. At $\sim$0.4 eV these oscillations are to a large extent suppressed in the experimental results of Fig. 2(a), but nevertheless seen near the BVB edge. However, in the experiment, due to the direct overlap between the TSS and BVB in this energy region, it is difficult to unambiguously disentangle the bulk from the surface contribution.

In conclusion, by utilizing tr-ARPES we probe the fully-symmetric longitudinal-optical coherent phonon A$_{1g}$ mode in the prototypical TI \BiTe. We resolve an anti-phase behavior in the onset of the coherent oscillation that is linked to a momentum-dependent coupling between phonons and TSS electrons in the presence of warping. Our {\it ab initio} calculations reproduce the experimental observation of a phase shift that is consistent with the divergence of the TSS from linearity. These results not only unravel, to the best of our knowledge, an unexpected behavior of the TSS in \BiTe\ but also demonstrate the capability of tr-ARPES to probe even more subtle details of the interplay between electron and collective modes in solid materials at ultimate timescales.

%\clearpage

\bibliographystyle{aipnum4-1}

\end{document}